\documentclass[two column,letter]{revtex4-1}
\usepackage{graphicx}
\begin{document}

\title{Local 3$d$ Electronic Structures of Co-Based Complexes \\
with Medicinal Molecules Probed by Soft X-Ray Absorption}
\author{Kohei Yamagami$^1$, Hidenori Fujiwara$^1$, Shin Imada$^2$, Toshiharu Kadono$^2$, Keisuke Yamanaka$^3$, Takayuki Muro$^4$, Arata Tanaka$^5$, Takuma Itai$^6$, Nobuto Yoshinari$^6$, Takumi Konno$^6$, and Akira Sekiyama$^1$}
\affiliation{$^1$Division of Materials Physics, Graduate School of Engineering Science, Osaka University, Toyonaka, Osaka 560-8531, Japan}
\affiliation{$^2$Department of Physical Science, Ritsumeikan University, Kusatsu, Shiga 525-8577, Japan}
\affiliation{$^3$Synchrotron Radiation Center, Ritsumeikan University, Kusatsu, Shiga 525-8577, Japan}
\affiliation{$^4$Japan Synchrotron Radiation Research Institute, Sayo, Hyogo 679-5198, Japan}
\affiliation{$^5$Department of Quantum Matter, ADSM, Hiroshima University, Higashihiroshima, Hiroshima 739-8530, Japan}
\affiliation{$^6$Department of Chemistry, Graduate School of Science, Osaka University, Toyonaka, Osaka 560-0043, Japan}

\begin{abstract}
We have examined the local 3$d$ electronic structures of Co-Au multinuclear complexes with the medicinal molecules {\scshape d}-penicillaminate ({\scshape d}-pen) [Co\{Au(PPh$_{3}$)({\scshape d}-pen)\}$_{2}$]ClO$_{4}$ and [Co$_{3}$\{Au$_{3}$(tdme)({\scshape d}-pen)$_{3}$\}$_{2}$] by Co $L_{2,3}$-edge soft X-ray absorption (XAS) spectroscopy, where PPh$_{3}$ denotes triphenylphosphine and tdme stands for 1,1,1-tris[(diphenylphosphino)methyl]ethane.
The Co $L_{{2,3}}$-edge XAS spectra indicate the localized ionic 3$d$ electronic states in both materials. 
The experimental spectra are well explained by spectral simulation for a localized Co ion under ligand fields with the full multiplet theory, which verifies that the ions are in the low-spin Co$^{3+}$ state in the former compound and in the high-spin Co$^{2+}$ state in the latter.
\end{abstract}

\maketitle
\section{Introduction}
\begin{figure}
\begin{center}
\includegraphics[width=8cm]{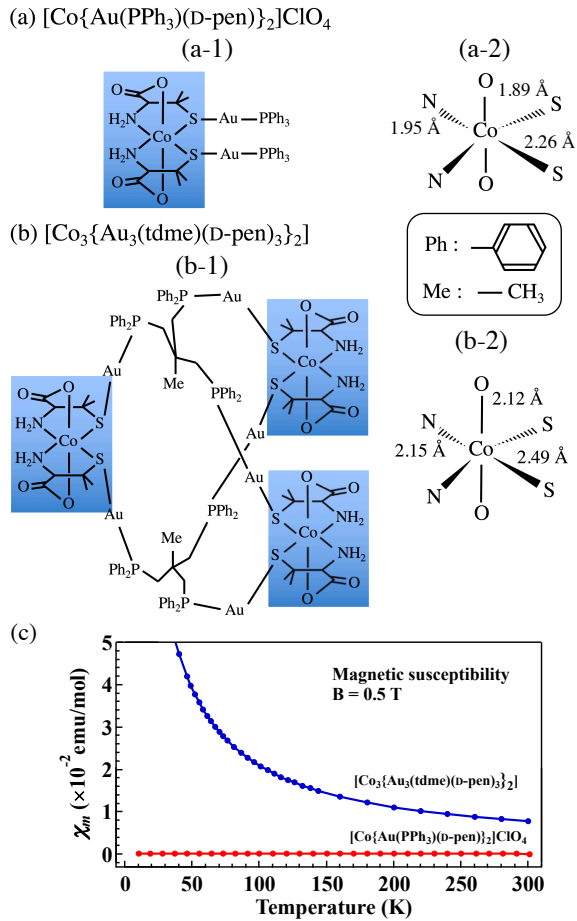}
\end{center}
\vspace{-5mm}
\begin{center}
\caption{(Color online) Molecular structures of (a) [Co\{Au(PPh$_{3}$)({\scshape d}-pen)\}$_{2}$]ClO$_{4}$, and (b) [Co$_{3}$\{Au$_{3}$(tdme)({\scshape d}-pen)$_{3}$\}$_{2}$]. Ph and Me denote the phenyl and methyl groups, respectively. (a-2) and (b-2) show the local structure around the Co ions and the bond lengths of Co-O, Co-N, and Co-S estimated from the X-ray diffraction. These complexes have the ligand environment with octahedral symmetry, displayed as blue marks.\cite{TMC[1],TMC[2]}(c) Temperature dependence of the magnetic susceptibility $\chi_{m}$ of [Co\{Au(PPh$_{3}$)({\scshape d}-pen)\}$_{2}$]ClO$_{4}$ and [Co$_{3}$\{Au$_{3}$(tdme)({\scshape d}-pen)$_{3}$\}$_{2}$].}
\label{TMC}
\end{center}
\end{figure}
\begin{figure}
\begin{center}
\includegraphics[width=8cm]{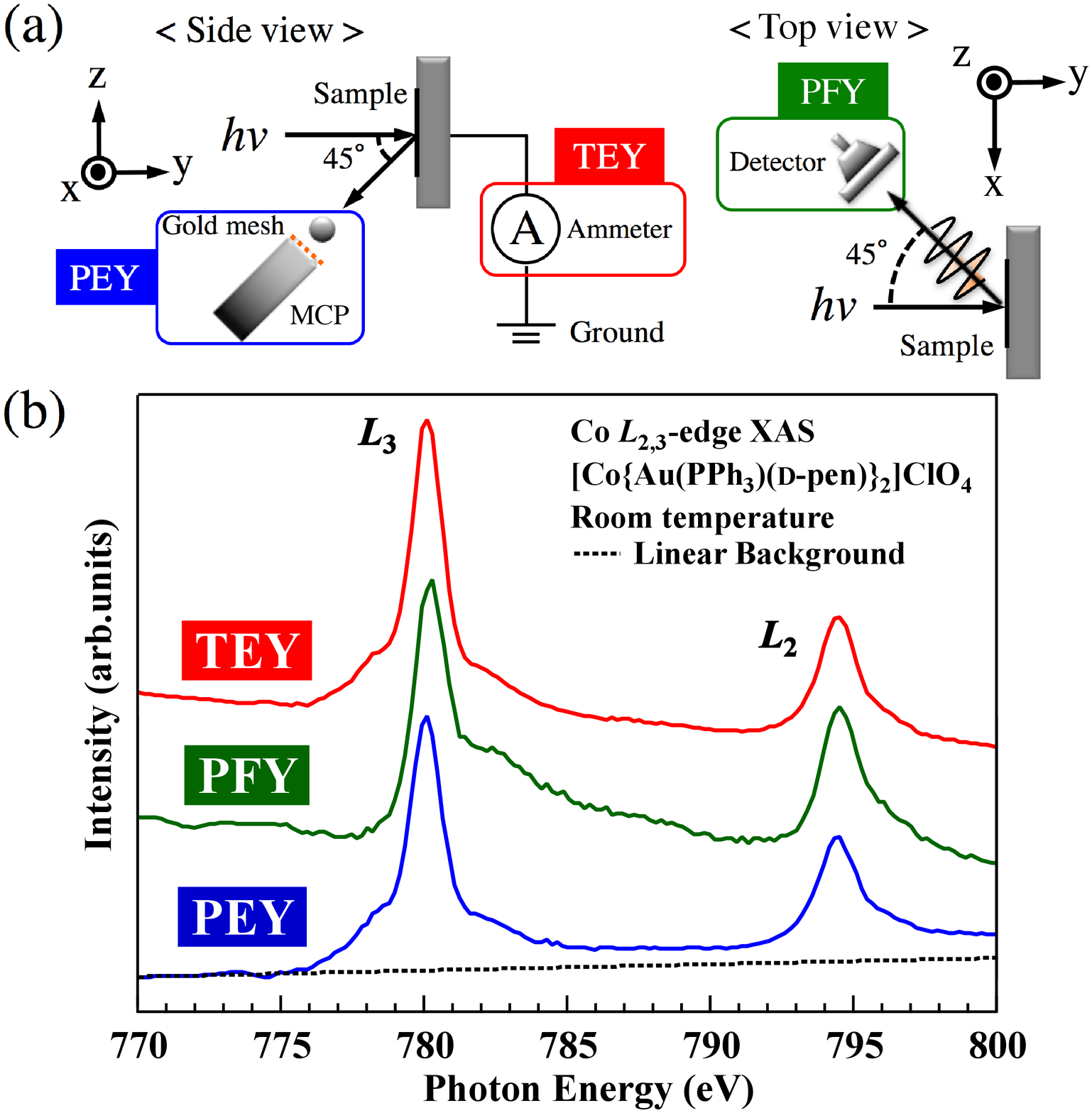}
\end{center}
\vspace{-5mm}
\begin{center}
\caption{(Color online) (a) Experimental geometry in the total electron yield (TEY), partial fluorescence yield (PFY), and partial electron yield (PEY) modes. (b) Co $L_{2,3}$-edge XAS raw spectra of [Co\{Au(PPh$_{3}$)({\scshape d}-pen)\}$_{2}$]ClO$_{4}$ in the TEY (red solid line), PFY (green solid line), and PEY (blue solid line) modes. The linear background of the spectrum in the PEY mode is shown by a dotted line. The linear background for the spectrum in the PEY mode was determined from the raw spectral weight in the region of $h\nu$ = 765$-$772 eV.}
\label{PEYTEYPFY}
\end{center}
\end{figure}
The electronic structure of a transition metal ion in materials is important for their functions not only in various inorganic crystalline solids but also in complexes with organic molecules as seen in hemoglobin.
Actually, many transition metal complexes with interesting natures such as spin crossover~\cite{SCO1, SCO2, SOC3} and mixed valence~\cite{MIXCOM1} have been synthesized.
Meanwhile, {\scshape d}-penicillaminate ({\scshape d}-pen) is employed as a heavy-metal antagonist~\cite{DPENI, DPENII, DPENIII} since it has the abillity to build metal coordination systems.
Such an aminothiolate is also a candidate raw material for newly functional metal multinuclear coordination systems.
So far, many metal complexes with {\scshape d}-pen have been developed~\cite{MIXCOM1, DPENCOMI, DPENCOMII}.
Among them, it has been reported that the Co$^{2+}$ ions in single-crystalline [Co$_{3}$\{Au$_{3}$(tdme)({\scshape d}-pen)$_{3}$\}$_{2}$] (tdme denotes 1,1,1-tris[(diphenylphosphino)methyl]ethane) are unusually stable in an octahedral structure bonded by the coordinations with two aliphatic thiolato, two amine, and two carboxylate donors~\cite{TMC[1]}, as shown in Fig. 1.
On the other hand, the Co ions in another Co-Au trinuclear complex [Co\{Au(PPh$_{3}$)({\scshape d}-pen)\}$_{2}$]ClO$_{4}$ (PPh$_{3}$ stands for triphenylphosphine) have been reported to be in the Co$^{3+}$ state, although the local coordination environment is similar to that in [Co$_{3}$\{Au$_{3}$(tdme)({\scshape d}-pen)$_{3}$\}$_{2}$] except for their bond length~\cite{TMC[2]}, as shown in Fig. 1.
It is well known that the Au$^{+}$ (5$d^{10}$) states are stabilized in the linear coordination~\cite{Aucoordination}, which is the case for [Co$_{3}$\{Au$_{3}$(tdme)({\scshape d}-pen)$_{3}$\}$_{2}$] and [Co\{Au(PPh$_{3}$)({\scshape d}-pen)\}$_{2}$]ClO$_{4}$.
On the other hand, the valence and spin states of the Co ions are not straightforwardly determined, as recognized for the inorganic oxide La$_{1.5}$Sr$_{0.5}$CoO$_{4}$, where highly localized high-spin (HS) Co$^{2+}$ and low-spin (LS) Co$^{3+}$ ions coexist within the same CoO$_{2}$ plane~\cite{REFCOMIII}.
For the Co-Au multinuclear complexes, they have so far been estimated from their crystal color, magnetic susceptibility, and light absorption in the visible and UV regions~\cite{TMC[1], TMC[2]}.
In particular, the temperature dependence of the magnetic susceptibility corrected by the Pascal method~\cite{PASCAL} [Fig. 1(c)] implies that Co$^{2+}$ ions in [Co$_{3}$\{Au$_{3}$(tdme)({\scshape d}-pen)$_{3}$\}$_{2}$] are in the paramagnetic high-spin (HS) state ($S=3/2$), whereas Co$^{3+}$ ions in [Co\{Au(PPh$_{3}$)({\scshape d}-pen)\}$_{2}$]ClO$_{4}$ are in the nonmagnetic low-spin (LS) state ($S=0$) by comparison with the calculated values~\cite{SPINONLYEQUATION}.
However, direct verifications using an element-specific probe are still lacking.
In addition, such fundamental information on the local electronic states around the Co ions, such as crystal-field splitting and the degree of hybridization effects (whether the charge-transfer effects from the ligand sites should be explicitly taken into account or not), is unclear for these complexes. \\
\ \ In the high-energy spectroscopic field, $K$-edge X-ray absorption fine structure (XAFS) has been applied to transition metal complexes~\cite{XAFS1, XAFS2, XAFS3}.
However, it is difficult to obtain the local 3$d$ electronic structure by XAFS as well as the $K$-edge X-ray absorption near-edge structure (XANES) due to the $s$-to-$p$ excitation process.
On the other hand, $L_{2,3}$-edge soft X-ray absorption (XAS) spectroscopy is powerful for directly probing the local element-specific electronic structure,~\cite{XASI,XASII,XASIII,XASIV,XASV,XASVI,Depth} as seen for La$_{1.5}$Sr$_{0.5}$CoO$_{4}$~\cite{REFCOMIII}.
However, the application of $L_{2,3}$-edge XAS to transition metal complexes is still a new frontier because sample degradation caused by soft X-ray irradiations has often prevented the acquisition of reliable spectra~\cite{RDamageI,RDamageII}.
These technical difficulties are solved by optimizing the photon density and appropriately changing the sample position.\\
\ \ In this paper, we report on the local 3$d$ electronic structures of the Co-Au multinuclear complexes [Co\{Au(PPh$_{3}$)({\scshape d}-pen)\}$_{2}$]ClO$_{4}$ and [Co$_{3}$\{Au$_{3}$(tdme)({\scshape d}-pen)$_{3}$\}$_{2}$] investigated by the Co $L_{2,3}$-edge XAS.
It has been verified that the Co ions are in the LS-Co$^{3+}$ state for the former compound, whereas they are in the HS-Co$^{2+}$ state for the latter system on the basis of the comparisons of the experimental spectra with spectral simulations for a localized ion under the crystalline electronic field (CEF) with taking the full multiplet theory into account.
\begin{figure}
\begin{center}
\includegraphics[width=8cm]{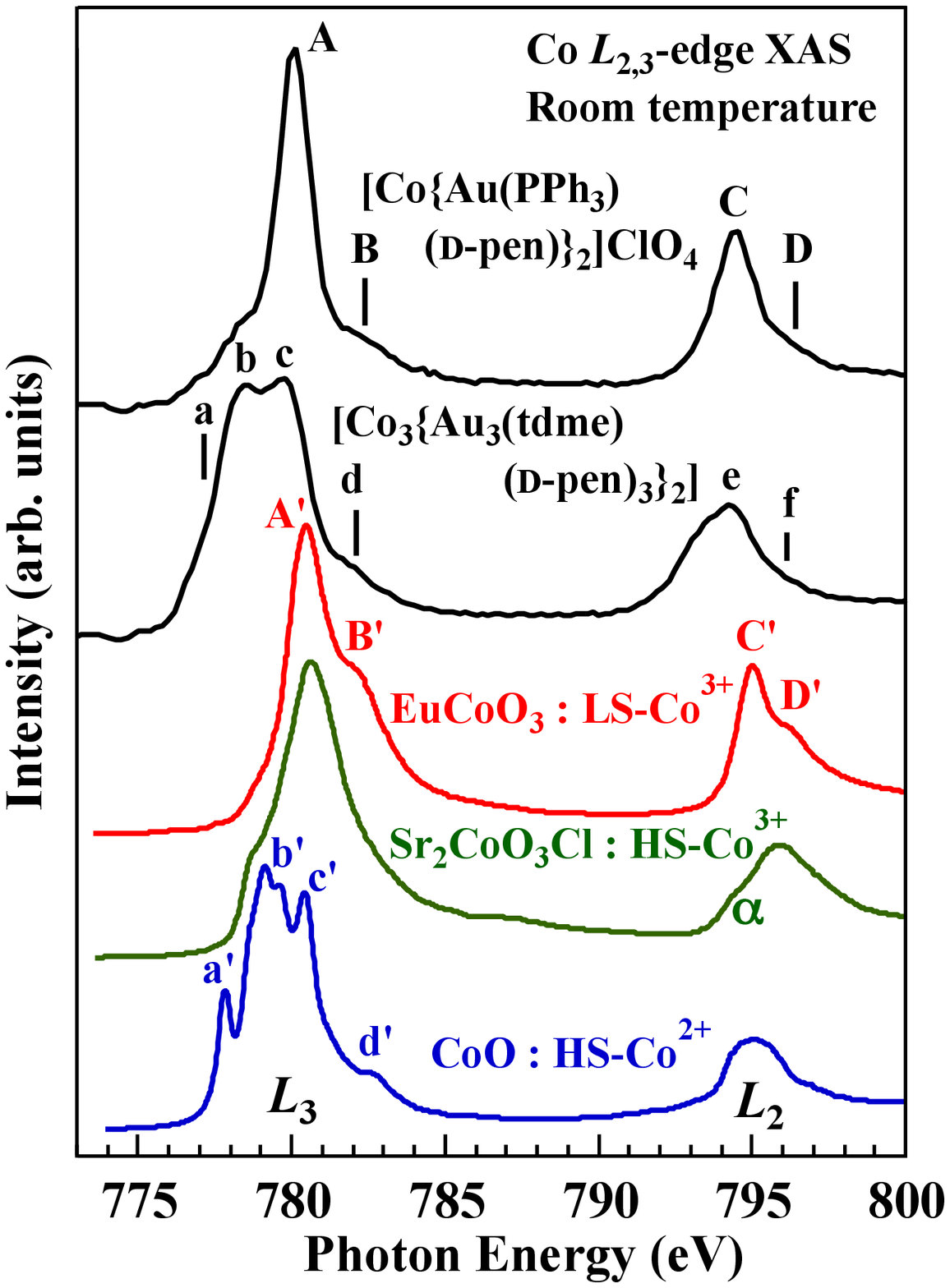}
\end{center}
\vspace{-5mm}
\begin{center}
\caption{(Color online) Co $L_{2,3}$-edge XAS spectra in the PEY mode for [Co\{Au(PPh$_{3}$)({\scshape d}-pen)\}$_{2}$]ClO$_{4}$ and [Co$_{3}$\{Au$_{3}$(tdme)({\scshape d}-pen)$_{3}$\}$_{2}$] (black solid line) compared with those reported for CoO (HS-Co$^{2+}$: blue solid line), Sr$_{2}$CoO$_{3}$Cl (HS-Co$^{3+}$: green solid line), and EuCoO$_{3}$(LS-Co$^{3+}$: red solid line) cited from Refs.13, 34, and 35. The characteristic spectral structures are labeled as A-D and a-f in our experimental spectra and as A'-D' and a'-d', $\alpha$ in the reference spectra.}
\label{ExperimentResults}
\end{center}
\end{figure}
\section{Experimental}
The XAS measurements were carried out at BL-11 of Synchrotron Radiation Center in Ritsumeikan University, Japan.
In this beamline, so-called varied-line-spacing plane gratings were employed, supplying monochromatic photons with $h\nu$ = 40$-$1000 eV.
The Co $L_{2,3}$-edge XAS spectra ($h\nu$ = 760$-$810 eV) were taken simultaneously in the total electron yield (TEY), partial fluorescence yield (PFY), and partial electron yield (PEY) modes with a photon energy resolution of $\sim$300 meV.
The experimental geometry is shown in Fig. ~\ref{PEYTEYPFY}(a).
In the PFY mode, a large-area silicon drift detector capturing photon-energy-dependent luminescence was set to 45$^{\circ}$ with respect to the photon propagation in the xy plane.
For the Co $L_{2,3}$-edge XAS measurements, the luminescence with $h\nu = 700-950$ eV including the Co $L_{2,3}$ lines was detected as a signal.
On the other hand, in the PEY mode, the microchannel plate (MCP) detecting the Auger and secondary electrons was set at an angle of 45$^{\circ}$ below the direction of photon propagation.
In the front of MCP, a gold mesh was installed to be able to apply a voltage.
We applied a voltage of $-$550 V to the mesh for the Co $L_{2,3}$-edge measurements to suppress the strong background caused by the C, N, and O $K$-edge absorptions in the spectra.\\
\ [Co\{Au(PPh$_{3}$)({\scshape d}-pen)\}$_{2}$]ClO$_{4}$ and [Co$_{3}$\{Au$_{3}$(tdme)({\scshape d}-pen)$_{3}$\}$_{2}$] were synthesized using the precursors and Co(CH$_{3}$COO)$_{2}$$\cdot$4H$_{2}$O in water.
The details of the synthesis have been reported in Refs. 10, and 11.
These powderlike single-crystalline samples were thinly expanded on the conductive carbon tape attached to the sample holder in air before transferring them to the vacuum chamber.
Since the size of every single-crystal piece was smaller than $\sim$100 $\mu$m, neither cleaving (fracturing) nor scraping of the samples in situ to obtain the clean surfaces was feasible for the XAS measurements.
However, we consider that the intrinsic XAS spectra of these samples were obtained, as discussed later, probably owing to the stability of the ionized Co states in the compounds, in which the Co ions have already been ``oxidized" in a chemical sense.
We repeatedly measured the spectra on the same and different sample positions, confirming the data reproducibility with neither serious radiation damage nor sample-position dependence of the Co $L_{2,3}$-edge XAS spectra.
The measurements were performed at room temperature. 
Preliminary $L_{2,3}$-edge XAS experiments were performed at BL27SU in SPring-8 to determine the feasibility of the method.\\
\section{Results and Discussions}
The raw Co $L_{2,3}$-edge XAS spectra for [Co\{Au(PPh$_{3}$)({\scshape d}-pen)\}$_{2}$]ClO$_{4}$ in the TEY, PFY, and PEY modes are shown in Fig.~\ref{PEYTEYPFY}(b).
There is a double-peak structure of the $L_{3}$ ($h\nu$ $\approx$ 780 eV) and $L_{2}$ ($h\nu$ $\approx$ 794 eV) edges due to the Co 2$p$ core-hole spin-orbit coupling in all spectra.
The overall spectral features are mutually consistent among these three spectra without showing a serious peak shift, indicating that we have successfully obtained the intrinsic XAS spectra.
On the other hand, there are some minor discrepancies among the spectra.
The shoulder structure at $h\nu$ = 779 eV in the spectra in the TEY and PEY modes is markedly suppressed in the PFY spectrum.
On the basis of the probing depth of each mode \cite{Depth}, we can judge that this shoulder is due to the surface contribution deviating from the bulk ones.
Between the two main $L_{2,3}$-edge peaks, another shoulder is seen at $h\nu$ = 782 eV in both the TEY and PEY spectra, which is overlapped with an asymmetric tail in the PFY spectrum.
Such an asymmetric tail is also seen on the high-$h\nu$ side of the $L_{2}$ edge in the PFY spectrum.
These tails are considered to originate from so-called self-absorption effects~\cite{Depth,PFYdiscussionI,PFYdiscussionII}, by which the spectral shape can deviate from the intrinsic one.
In addition, the background of the spectrum in the TEY mode depends on $h\nu$, where it is stronger for the low-$h\nu$ side than for the high-$h\nu$ side.
This suggests that the intrinsic Co $L_{2,3}$-edge XAS spectral weight is markedly weaker than the background yielded by high-energy tails of the C, N, and O $K$-edge absorptions caused by a small Co element ratio as C:N:O:Co = 46:2:8:1 for [Co\{Au(PPh$_{3}$)({\scshape d}-pen)\}$_{2}$]ClO$_{4}$.
In the PEY mode, on the other hand, the background is relatively weaker than the Co $L_{2,3}$-edge XAS spectral weight owing to the suppression of the contribution of electrons with kinetic energy less than 550 eV to the spectral intensity.
This background increases slightly and linearly with $h\nu$, as shown in Fig.~\ref{PEYTEYPFY}(b).
Therefore, we have concluded that the spectrum in the PEY mode is the most reliable for the quantitative discussion shown below even though it is relatively surface-sensitive.\\
\begin{table}[h]
  \caption{Values of the Slater integrals ($F$, $G$), and spin-orbit couplings ($\zeta$) in units of eV used in the various Co configurations for the theoretical calculations.}
  \begin{tabular}{ccccc}
\\
   \hline
   \hline
             & 2$p^{6}$3$d^{6}$ & 2$p^{5}$3$d^{7}$ & 2$p^{6}$3$d^{7}$ & 2$p^{5}$3$d^{8}$ \\
    \hline
    $F^{2}(3d3d)$ & 10.138 & 10.743 & 9.295 & 9.925 \\
    $F^{4}(3d3d)$ & 6.339 & 6.720 & 5.775& 6.172 \\
    $F^{2}(2p3d)$ & $-$ & 6.321 & $-$ & 5.810 \\
    $G^{1}(2p3d)$ & $-$ & 4.760 & $-$ & 4.317 \\
    $G^{3}(2p3d)$ & $-$ & 2.709 & $-$ & 2.455 \\
    $\zeta_{3d}$ & 0.074 & 0.092 & 0.066 & 0.083 \\
    $\zeta_{2p}$ & $-$ & 9.742 & $-$ & 9.744 \\
    \hline
    \hline
  \end{tabular}
\end{table}
\ \ The background-subtracted Co $L_{2,3}$-edge XAS spectra of [Co\{Au(PPh$_{3}$)({\scshape d}-pen)\}$_{2}$]ClO$_{4}$ and [Co$_{3}$\{Au$_{3}$(tdme)({\scshape d}-pen)$_{3}$\}$_{2}$]  in the PEY mode are shown in Fig. 3.
The characteristic spectral structures originating from the bulk nature are labeled as A-D and a-f in the spectra, respectively.
A metal-like asymmetric tail of the $L_{3}$ main peak~\cite{Depth,YAMASAKI} is not seen.
A possible satellite structure in the photon energy range of $5-6$ eV higher than the main peaks, called a charge-transfer satellite, is negligible.
Such a satellite originates from hybridizations between the Co 3$d$ and ligand $p$ orbitals described by configuration interactions on the basis of the cluster model~\cite{SateI,SateII} or the Anderson impurity model~\cite{AnderI}. 
Therefore, the absence of the satellite suggests that these spectra can be well explained by local ion models without explicitly considering the hybridization effects.
Let us compare our spectra with those reported for the inorganic Co compounds EuCoO$_{3}$ (pure LS-Co$^{3+}$), Sr$_{2}$CoO$_{3}$Cl (pure HS-Co$^{3+}$), and CoO (pure HS-Co$^{2+}$)~\cite{REFCOMI,REFCOMII,REFCOMIII} indicated in Fig. 3, where the characteristic spectral structures are also labeled as A'-D', a'-f', and $\alpha$, respectively.
The spectral features of [Co\{Au(PPh$_{3}$)({\scshape d}-pen)\}$_{2}$]ClO$_{4}$ are inconsistent with those of Sr$_{2}$CoO$_{3}$Cl near the $L_{2}$ edge, while structures A-D seem to correspond well to structures A'-D' for EuCoO$_{3}$.
Therefore, it is concluded that the Co ions of [Co\{Au(PPh$_{3}$)({\scshape d}-pen)\}$_{2}$]ClO$_{4}$ are in the LS-Co$^{3+}$ state.
On the other hand, structures a-d in the spectrum of [Co$_{3}$\{Au$_{3}$(tdme)({\scshape d}-pen)$_{3}$\}$_{2}$] are similar to structures a'-d' in the spectrum of CoO at the $L_{3}$ edge although they are somehow smeared.
The spectral shape of [Co$_{3}$\{Au$_{3}$(tdme)({\scshape d}-pen)$_{3}$\}$_{2}$] at the $L_{2}$ edge seems to be rather different from that of CoO, but the $L_{3}$-$L_{2}$ splitting energy is mutually consistent.
We can thus judge that the Co ions in [Co$_{3}$\{Au$_{3}$(tdme)({\scshape d}-pen)$_{3}$\}$_{2}$] are in the HS-Co$^{2+}$ state.
We consider that the shoulder structure at $h\nu$ = 779 eV in the spectrum of the [Co\{Au(PPh$_{3}$)({\scshape d}-pen)\}$_{2}$]ClO$_{4}$ is caused by the surface Co$^{2+}$ contribution corresponding to peak b for [Co$_{3}$\{Au$_{3}$(tdme)({\scshape d}-pen)$_{3}$\}$_{2}$], where the Co$^{2+}$ state is favored in the surface due to its larger ionic radius than that for the Co$^{3+}$ state.\\
\begin{figure}
\begin{center}
\includegraphics[width=8cm]{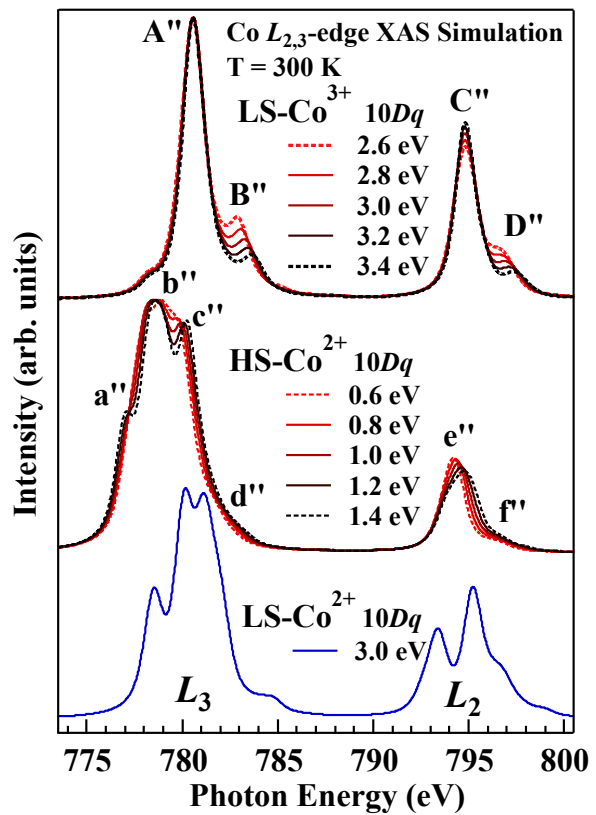}
\end{center}
\vspace{-5mm}
\begin{center}
\caption{(Color online) Simulated Co $L_{2,3}$-edge XAS spectra with the full-multiplet theory for Co$^{3+}$ and Co$^{2+}$ ions as a function of the CEF splitting 10$Dq$ in the $O_{h}$ symmetry. Gaussian and Lorentzian broadenings were set to $\sim$300 and $\sim$400 meV, respectively, to reproduce the XAS spectra. The characteristic structures are labeled as A"-D" and a"-f" in the spectra of Co$^{3+}$ and Co$^{2+}$, respectively.}
\label{Xtlscalculation}
\end{center}
\end{figure}
\ \ As mentioned earlier, [Co\{Au(PPh$_{3}$)({\scshape d}-pen)\}$_{2}$]ClO$_{4}$ and [Co$_{3}$\{Au$_{3}$(tdme)({\scshape d}-pen)$_{3}$\}$_{2}$] have an octahedral coordination environment.
The average bond lengths of the Co$^{3+}$ system [Co\{Au(PPh$_{3}$)({\scshape d}-pen)\}$_{2}$]ClO$_{4}$ has been estimated as 2.03 \AA~\cite{TMC[2]}, which is larger than that for EuCoO$_{3}$ (1.93 \AA)~\cite{ECOlattice}. 
For the Co$^{2+}$ systems, the average Co bond length of [Co$_{3}$\{Au$_{3}$(tdme)({\scshape d}-pen)$_{3}$\}$_{2}$] (2.25 \AA)~\cite{TMC[1]} is also larger than that of CoO (2.10 \AA)~\cite{COlattice}.
The larger bond lengths for the complexes than for the inorganic oxides with the same valence might allow us to discuss the Co $L_{2,3}$-edge XAS spectral shape in view of the localized ions under CEF. 
Since it is well known that the local HS (LS) state of transition metal ions is stabilized by lower (larger) effective CEF splitting, we performed spectral simulations for the ground states of the Co ion~\cite{FullMultiplet, Function10Dq} in cubic $O_{h}$ symmetry using the XTLS 9.0 program~\cite{Xtls}.
All atomic parameters such as the 3$d$-3$d$ and 2$p$-3$d$ coulomb and exchange interactions (Slater integrals $F$ and $G$), and the 2$p$ and 3$d$ spin-orbit couplings were obtained using Cowan's code~\cite{Cowan} based on the Hartree$-$Fock method.
The Slater integrals were reduced to 80$\%$~\cite{XASV,CROSSOVER} to reproduce the Co $L_{2,3}$-edge XAS spectra, as shown in Table I.
The actual local symmetry of the Co ions is lower than the cubic ones for both materials.
However, the CEF splitting 10$Dq$ between the $t_{2g}$ and $e_{g}$ states in the $O_{h}$ symmetry is crucial for determining the local spin state (HS or LS) of the Co ions.
Therefore, the simulations in the $O_{h}$ symmetry should give meaningful information on the 3$d$ electronic states.
The simulation results for Co$^{3+}$ and Co$^{2+}$ states as a function of 10$Dq$ in the $O_{h}$ symmetry are shown in Fig.~\ref{Xtlscalculation} and are consistent with previous calculation results~\cite{Function10Dq}.
Structures A"-D" and a"-f" denote the characteristic structures for the Co ions.
For the spectra of Co$^{3+}$ states, the shoulder structure B" (D") is reduced and shifted to a higher photon energy with increasing 10$Dq$.
As well as structures B" and D", structures c" and f" are shifted to a higher photon energy in the spectra of Co$^{2+}$ states.
In contrast, structure a" is shifted to the low-$h\nu$ side with increasing $10Dq$.
Meanwhile, the spectrum of LS-Co$^{2+}$ states with $10Dq$ = 3.0 eV is completely inconsistent with the XAS results for  [Co$_{3}$\{Au$_{3}$(tdme)({\scshape d}-pen)$_{3}$\}$_{2}$] since the $L_{2}$-edge main-peak structure is clearly different.\\
\ \ Our results show that the energy difference between A (b) and B (c) of [Co\{Au(PPh$_{3}$)({\scshape d}-pen)\}$_{2}$]ClO$_{4}$ ([Co$_{3}$\{Au$_{3}$(tdme)({\scshape d}-pen)$_{3}$\}$_{2}$]) is 2.28 eV (1.22 eV).
Figure 5 shows a comparison of the experimental spectra with the best simulated ones by optimizing 10$Dq$ so as to reproduce the energy difference.
From these analyses, we determined the 10$Dq$ values of 3.0$\pm$0.2 eV for [Co\{Au(PPh$_{3}$)({\scshape d}-pen)\}$_{2}$]ClO$_{4}$ and 1.0$\pm$0.2 eV for [Co$_{3}$\{Au$_{3}$(tdme)({\scshape d}-pen)$_{3}$\}$_{2}$].
These 10$Dq$ values lead to the LS (HS) state for the former (latter) compound, consistent with the magnetic susceptibility~\cite{SPINONLYEQUATION}  and the optical absorption results.
The slight quantitative inconsistency of the line shape at the $L_{3}$ edge between [Co$_{3}$\{Au$_{3}$(tdme)({\scshape d}-pen)$_{3}$\}$_{2}$] and the HS-Co$^{2+}$ state might be caused by the difference in symmetry.\\
\begin{figure}
\begin{center}
\includegraphics[width=8cm]{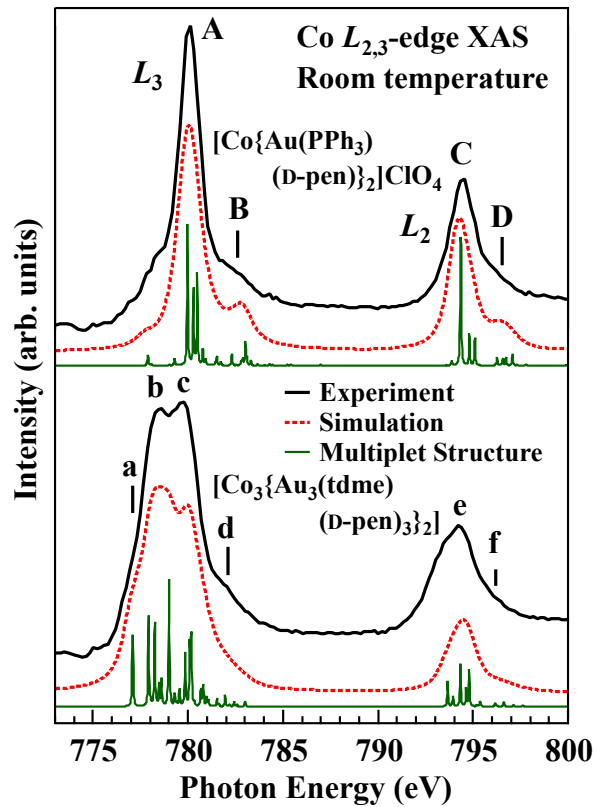}
\end{center}
\vspace{-5mm}
\begin{center}
\caption{(Color online) Comparison of experimental (black solid lines) and theoretically reproduced (red solid lines) spectra for [Co\{Au(PPh$_{3}$)({\scshape d}-pen)\}$_{2}$]ClO$_{4}$ and [Co$_{3}$\{Au$_{3}$(tdme)({\scshape d}-pen)$_{3}$\}$_{2}$] with 10$Dq$ = 3.0 and 1.0 eV, respectively. The XAS final-state multiplet structures (green solid line) are also shown.}
\label{Optimization}
\end{center}
\end{figure}
\ \ Note that our optimized 10$Dq$ value of 3.0 eV for the LS-Co$^{3+}$ states is much (more than 3 times) larger than that for EuCoO$_{3}$~\cite{REFCOMII} since our value corresponds to a $renormalized$ 10$Dq$ in which the hybridization effects (anisotropy in the hybridizations in a rigorous sense) are implicitly included.
This is higher than the value of the LS-HS crossover ($\approx$ 2.2 eV)~\cite{CROSSOVER}.
The estimated 10$Dq$ of 1.0 eV for the HS-Co$^{2+}$ state is expected to be much smaller than that of the LS-HS crossover for Co$^{2+}$ ($\sim$ 2.0$-$2.5 eV).
Such a small 10$Dq$ value might unusually stabilize the Co$^{2+}$ state in [Co$_{3}$\{Au$_{3}$(tdme)({\scshape d}-pen)$_{3}$\}$_{2}$]. 
When we consider the fact that the Co$^{3+}$ ions have a smaller ionic radius than the Co$^{2+}$ ions in these compounds, the larger 10$Dq$ in [Co\{Au(PPh$_{3}$)({\scshape d}-pen)\}$_{2}$]ClO$_{4}$ is understood as a consequence of the  relatively larger hybridization strength caused by the shorter distance between the Co and the neighboring ions than that for [Co$_{3}$\{Au$_{3}$(tdme)({\scshape d}-pen)$_{3}$\}$_{2}$].
Spectral simulations also lead to spin moments of 1.87 for Co$^{2+}$ states and 0.0 for Co$^{3+}$ states, consistent with those obtained from the magnetic susceptibility at room temperature~\cite{SPINONLYEQUATION}.
In addition, the Tanabe$-$Sugano diagram~\cite{diagram} for Co ions can explain the optical absorption results~\cite{TMC[1],TMC[2]} based on 10$Dq$ obtained from our results, which indicates that the intrinsic Co 3$d$ electronic states of Co multinuclear complexes have been successfully identified.
\section{Conclusions}
We have studied the 3$d$ electronic structure of Co-Au multinuclear complexes with the medicinal molecules {\scshape d}-penicillaminate [Co\{Au(PPh$_{3}$)({\scshape d}-pen)\}$_{2}$]ClO$_{4}$ and [Co$_{3}$\{Au$_{3}$(tdme)({\scshape d}-pen)$_{3}$\}$_{2}$] by Co $L_{2,3}$-edge XAS.
There was no observation of the satellite structure in [Co\{Au(PPh$_{3}$)({\scshape d}-pen)\}$_{2}$]ClO$_{4}$ and [Co$_{3}$\{Au$_{3}$(tdme)({\scshape d}-pen)$_{3}$\}$_{2}$], indicating that the Co ions are strongly localized.
From the spectral calculations for the ions under CEF in the $O_{h}$ symmetry, we have verified that the Co ions are in the LS-Co$^{3+}$ state for [Co\{Au(PPh$_{3}$)({\scshape d}-pen)\}$_{2}$]ClO$_{4}$ and in the HS-Co$^{2+}$ state for [Co$_{3}$\{Au$_{3}$(tdme)({\scshape d}-pen)$_{3}$\}$_{2}$] with a much smaller 10$Dq$, which are consistent with the magnetic susceptibility and optical absorption results.\\

\acknowledgments
We thank M. Murata, M. Yamada, and T. Ohta for supporting the experiments. 
The Soft X-ray XAS study was supported by the Project for the Creation of Research Platforms and Sharing of Advanced Research Infrastructure, Japan (No. R1511).
K. Yamagami was supported by the Program for Leading Graduate Schools ``Interactive Materials Science Cadet Program".
The results of this study benefitted considerably from pilot measurements performed at BL27SU in SPring-8 (2014B1299).


\begin{thebibliography}{99}
\bibitem{SPINONLYEQUATION}The magnetic susceptibility ($\chi_{m}$) in the paramagnetic phase of these materials~\cite{TMC[1],TMC[2]} follows the Curie law ($\chi_{m}=C/T$) above 100 K. 
In previous studies, $\chi_{m}T$ in units of emu$\cdot$K/mol at 300 K has been estimated as 2.38 for [Co$_{3}$\{Au$_{3}$(tdme)({\scshape d}-pen)$_{3}$\}$_{2}$] and 0.0 for [Co\{Au(PPh$_{3}$)({\scshape d}-pen)\}$_{2}$]ClO$_{4}$. These values correspond well to the calculated values of the effective magnetic moments of $2\sqrt{S(S+1)}=2.828\sqrt{\chi_{m}T}$ in which only spin components are taken into account due to the ligand field effects.
\end{thebibliography}
\end{document}